\documentclass{article}
\usepackage{graphicx}
\usepackage{amssymb}

\title{Classification of KPZQ and BDP models by \\ multiaffine analysis}

\author{\textsc{Hiroaki Katsuragi}\thanks{E-mail address: katsurag@asem.kyuhu-u.ac.jp} and \textsc{Haruo Honjo}
\\
\\
Department of Applied Science for Electronics and Materials, \\
Interdisciplinary Graduate School of Engineering Sciences, \\
Kyushu University, 6-1 Kasugakoen, Kasuga, Fukuoka 816-8580, Japan
\\}

\begin{document}

\maketitle

\begin{abstract}
We argue differences between the Kardar-Parisi-Zhang with Quenched disorder (KPZQ) and the Ballistic Deposition with Power-law noise (BDP) models, using the multiaffine analysis method. The KPZQ and the BDP models show {\it mono-affinity} and {\it multiaffinity}, respectively. This difference results from the different distribution types of neighbor-height differences in growth paths. Exponential and power-law distributions are observed in the KPZQ and the BDP, respectively. In addition, we point out the difference of profiles directly, i.e., although the surface profiles of both models and the growth path of the BDP model are rough, the growth path of the KPZQ model is smooth. 
\end{abstract}

\section{Introduction}
\label{}
The motion of rough surface has attracted much attention in recent decades \cite{Barabasi1,Family1}. The most important feature is a dynamic scaling of surface width (root mean square roughness) $w$. Consider a flat surface initially, and it grows to rough surface with time evolution. There are two scaling exponents $\alpha$ (roughness exponent) and $\beta$ (growth exponent). The $w$ is scaled by a scaling function $\Psi$ as $w\sim x^{\alpha} \Psi(t/x^z)$, where $x$, $t$, and $z$ are the space length scale, time scale, and dynamic exponent $z= \alpha / \beta$, respectively. The scaling function $\Psi$ consists of two parts depending on the argument $u=t/x^z$; $\Psi(u) \sim u^\beta$ for $u \ll 1$, and $\Psi(u) \sim const.$ for $u \gg 1$ \cite{Barabasi1,Family1}. 

Since Kardar, Parisi, and Zhang (KPZ) proposed the continuous partial differential equation model \cite{Kardar1}, many theoretical and simulational studies have been done. The KPZ equation is written as follows, 
\begin{equation}
 \partial_t h = \nu \nabla^2 h + \frac{\lambda}{2} (\nabla h)^2 + \eta(x,t). \label{eq:KPZ}
\end{equation}
Where, $h$, $\nu$, $\lambda$, and $\eta$ correspond to surface height, effective surface tension, lateral growth strength, and noise, respectively. They computed scaling exponents as $\alpha=1/2$ and $\beta=1/3$ for $1+1$ dimensional growing surfaces. Moreover, they found the scaling-law $\alpha + z =2$. This is a so-called KPZ universality scaling law. 

On the other hand, a lot of experiments of growing rough surface have been carried out. We show some typical results in Table \ref{tab:ExpResult}. As can be seen in Table \ref{tab:ExpResult}, all experiments show $\alpha >1/2$ for anomalous scaling (short range) regime. This implies the KPZ equation model is not a sufficient model for growing rough surface phenomena. Short range anomalous scaling has been studied in other previous papers \cite{Lopez1,Cshok1,Zhang2}. In ref.\ \ref{ref:Lopez1}, many models for kinetic rough surface were considered. However, noise was assumed as Gaussian white in all cases. In refs.\ \ref{ref:Cshok1} and \ref{ref:Zhang2}, they used noises different from Gaussian white. We focus on this short range anomalous scaling regime from the point of view of the relation between multiaffinity and noise statistics, in this paper. Therefore, we performed numerical simulations of two variants of the KPZ model according to refs.\ \ref{ref:Cshok1} and \ref{ref:Zhang2}.
\begin{table}
\begin{center}
\caption{Experimental or observed results of $\alpha$ in anomalous scaling short range regime. All results show $\alpha >1/2$.}
\begin{tabular}{lll}
\hline
Experiment & $\alpha$ & Reference \\
\hline \hline
Fluid flow    & 0.73 & Rubio {\it et al.}, 1989 \cite{Rubio1} \\
 & 0.81 & Horvath {\it et al.}, 1991 \cite{Horvath1} \\
 & 0.65-0.91        & He {\it et al.}, 1992 \cite{He1} \\
\hline
Paper wetting    & 0.63 &Buldyrev {\it et al.}, 1992 \cite{Buldyrev1} \\
\hline
Bacteria growth    & 0.78 & Vicsek {\it et al.}, 1990 \cite{Vicsek1} \\
 & 0.78 & Wakita {\it et al.}, 1997 \cite{Wakita1} \\
\hline
Burning front    & 0.78 & Zhang {\it et al.}, 1992 \cite{Zhang1} \\
 & 0.81-0.89 &Myllys {\it et al.}, 2000 \cite{Myllys1} \\
\hline
Crystal growth     & 0.79-0.91 & Honjo and Ohta, 1994 \cite{Honjo1} \\
\hline 
Mountains profile     & 0.73  & Matsushita and Ouchi,1989 \cite{Matsushita1} \\
 & 0.68  & Katsuragi and Honjo, 1999 \cite{Katsuragi1} \\
\hline
\label{tab:ExpResult}
\end{tabular}
\end{center}
\end{table}
In those models, the noise term is modified. In the one case, quenched disorder type noise is considered \cite{Cshok1}.  In the other case, uncorrelated power-law distributed noise is adopted \cite{Zhang2}. Both models can generate the large $\alpha$($>1/2$) values. However, physical meanings of those two models are quite different. How can we distinguish those two models? The answer we present in this paper is the multiaffine analysis method. 

In this paper, we report on the result of numerical simulations of two variants of the KPZ model. In the next section, multiaffine analysis method is described. We introduce details of two models in \S \ref{sec:Model}. And we apply multiaffine analysis method to simulations results, in \S \ref{sec:results}. In \S \ref{sec:discussion}, the roughness of growth path at each model is discussed. The comparison between our result and experimental data will be contained in this section. Furthermore, distribution forms of neighbor-height-differences are discussed, too. Finally, we conclude our results in the last section.

\section{Analysis methods}\label{sec:AM}
The multiaffine analysis method is based on the multifractal concept \cite{Barabasi1,Myllys1,Katsuragi1,Feder1}. According to their results, $q$-th order height-height correlation function $C_q$ is defined  as, 
\begin{equation}
C_q (x,t) = \langle |\delta h(x',t') - \delta h(x'+x,t'+t)|^q \rangle_{x',t'}. \label{eq:Cq(x,t)def}
\end{equation}
Where $\delta h(x,t)$ is a deviation from mean height, i.e., $\delta h(x,t) = h(x,t) - \langle h(t) \rangle_x$. If we fix the temporal (or spacial) coordinate, we can define the $q$-th order roughness exponent $\alpha_q$ and $q$-th order growth exponent $\beta_q$ at the limit of $x \to 0$ and $t \to 0$ as follows, 
{\setcounter{enumi}{\value{equation}}
\addtocounter{enumi}{1}
\setcounter{equation}{0}
\renewcommand{\theequation}{\theenumi\alph{equation}}
\begin{equation}
C_q(x,0) \sim x^{q \alpha_q}, \label{eq:aqdef}
\end{equation}
\begin{equation}
C_q(0,t) \sim t^{q \beta_q}.  \label{eq:bqdef}
\end{equation}
\setcounter{equation}{\value{enumi}}}
Obviously, $\alpha_q$ and $\beta_q$ with this definition quantify $q$-th order roughness scaling of surface profile $\delta h(x,t')$ (at fixed $t'$) and growth path $\delta h(x',t)$ (at fixed $x'$), respectively. If the $\alpha_q$ ($\beta_q$) varies with $q$, the surface profile (growth path) is multiaffine (multi-growth). If they show constant values independent of $q$, it means mono-affinity and mono-growth property. More details about multiaffinity can be found in refs.\ \ref{ref:Barabasi1}, \ref{ref:Katsuragi1} and \ref{ref:Katsuragi2}. The $\alpha_q$ ($\beta_q$) just corresponds to the generalized dimension $D_q$ in the multifractal formalism \cite{Feder1}. When we find the changing $D_q$ depending on $q$, we call that multifractal. Multiaffinity is an analogue of that. 

We can assume another definition of the growing exponent. That is based on the $q$-th order surface width $w_q$ defined as follows,
\begin{equation}
w_q = ( \frac{1}{L} \sum_{i=1}^{L} | \delta h_i |^q )^{1/q}. \label{eq:w_q}
\end{equation}
where $\delta h_i$, $L$ are height deviation at $i$-th site ($\delta h_i = h_i - \langle h_i \rangle$) and the number of sites, respectively. Then, the scaling forms can be written as $w_q \sim x^{\alpha_q}$ and $w_q \sim t^{\beta_q}$. 
We use this definition of $\beta_q$ later. 
Using those $C_q$ and $w_q$ analysis methods, we classify two models of growing rough surface and discuss its original mechanism.

\section{Models}\label{sec:Model}

\subsection{KPZQ model}\label{sec:KPZQ}
\begin{figure}
\begin{center}
\scalebox{0.7}[0.7]{\includegraphics{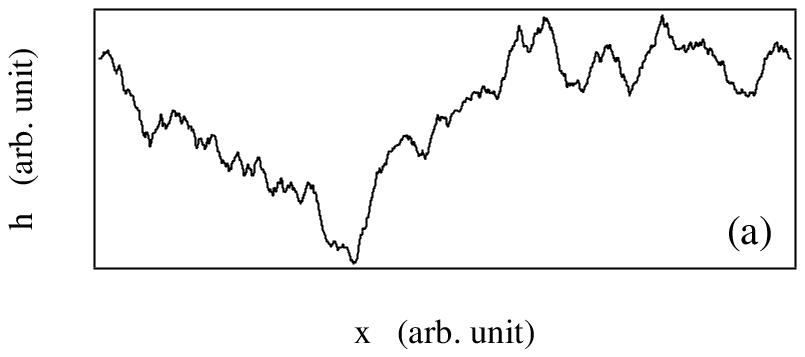}}
\scalebox{0.7}[0.7]{\includegraphics{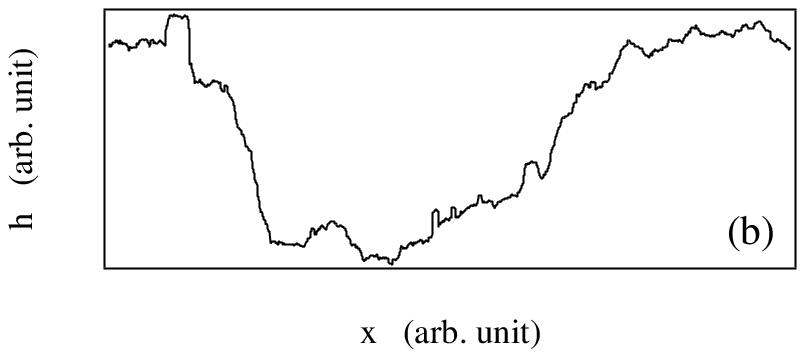}}
\caption{Typical surface profiles of (a) KPZQ model and (b) BDP model. The parameters are taken as system size $L=4096$ and calculation step $T=10^6$ steps in both models.}
\label{fig:SurfaceProfiles}
\end{center}
\end{figure}

We examine two variants of KPZ model. The one is the Kardar-Parisi-Zhang with Quenched disorder (KPZQ) model that employs quenched disorder instead of time dependent noise. This model can be written as follows by a transform of KPZ noise term $\eta(x,t) \;\to \; \eta(x,h)$,
\begin{equation}
 \partial_t h = \nu \nabla^2 h + \frac{\lambda}{2} (\nabla h)^2 + f + \eta(x,h). \label{eq:KPZQ}
\end{equation}
Where $f$ is a positive driving force term. If there is not any such positive driving force, growth of surface shall stop due to the effect of negative quenched noise. Csah\'ok {\it et al.}, calculated the characteristic exponents as $\alpha=0.75$ and $\beta=0.6$, by the dimensional analysis method \cite{Cshok1}. These exponents satisfy the KPZ universality scaling-law $\alpha + z =2$. The discretized version of eq.\ (\ref{eq:KPZQ}) with simple single step method is written as,
\begin{eqnarray}
h(x,t+\Delta t)&=&h(x,t) + \Delta t \Bigl( \{ h(x-1,t) - 2h(x,t) + h(x+1,t) \} \nonumber \\ 
&&+ \frac{\lambda}{2} \{ h(x+1,t) - h(x-1,t) \}^2 + v + \eta(x,[ h(x,t) ] ) \Bigr). \label{eq:discreteKPZQ}
\end{eqnarray}
Where $[ h(x,t) ]$ denotes integer part of $h(x,t)$. We use quenched disorder $\eta(x,h)$ with a uniform distributed noise $[-1,1]$.  The parameters $\lambda$, $f$, and $\Delta t$ are taken as $1.0$, $0.05$, and $0.01$ in all simulations. These values are close to pinning threshold. There is a crossover between short range anomalous scaling and asymptotic KPZ scaling, in this model \cite{Leschhorn1}. We show a surface profile example of the KPZQ model in Fig.\ \ref{fig:SurfaceProfiles}(a).   

\subsection{BDP model}
The other model is the Ballistic Deposition with Power-law noise (BDP) model. Consider a flat substrate. Then, the noise particles are dropped from above randomly. They stick growing aggregate when they reach its nearest neighbor and next-nearest neighbor. This growing rule corresponds to ultra discrete version of KPZ equation \cite{Nagatani1}. The rule is written as,
\begin{eqnarray}
h(x,t+1) &=& \max [h(x-1,t)+\eta(x-1,t), h(x,t) +\eta(x,t), \nonumber \\
&& h(x+1,t)+\eta(x+1,t)]. \label{eq:BDrule}
\end{eqnarray}
Zhang obtained large $\alpha$($>1/2$) value using the uncorrelated power-law distributed noise with this model \cite{Zhang2}.  Noise amplitude distributes as $P(\eta) \sim \frac{1}{\eta^{(\mu+1)}}$ for $\eta>1$ and $P(\eta)=0$ for otherwise. Barab\'asi {\it et al.}, found the multiaffinity in this model \cite{Barabasi2}. Buldyrev {\it et al.}, calculated the exponents as $\alpha = \frac{d+2}{\mu+1}$ and $\beta = \frac{d+2}{2\mu-d}$ with mean field idea \cite{Bul2} in the region $\mu<2d+3$. Where $d$ is a spatial dimension of substrate. In addition, We have revealed the asymptotic behavior of $\alpha_q$ and $\beta_q$ for large $q$ regime \cite{Katsuragi2}. Some experiments show the power-law distributed effective noise with $\mu \simeq 3$ \cite{Myllys1,Horvath2}. A surface profile example of this model at $\mu=3.0$ ($d=1$) is shown in Fig.\ \ref{fig:SurfaceProfiles}(b). In this case, $\alpha$ and $\beta$ become $0.75$ and $0.6$, respectively. These values coincide with those obtained from the KPZQ model as mentioned in \S \ref{sec:KPZQ}. Therefore, we cannot distinguish those two models only by the analysis of $\alpha$ and $\beta$. Thus we use multiaffine analysis henceforth.

\section{Simulations and results}
\label{sec:results}
\begin{figure}
\begin{center}
\scalebox{0.7}[0.7]{\includegraphics{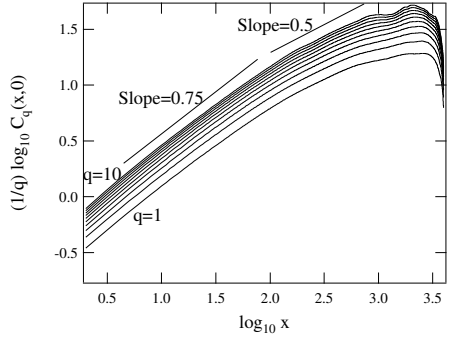}}
\caption{$\frac{1}{q}\log C_q(x,t')$ as a function of \ $\log x$ of the KPZQ model. Each curve has the same slope $0.75$ that indicates mono-affinity of the KPZQ model. Moreover, it is consistent with the dimensional analysis of Csah\'ok {\it et al.}. The parameters are taken as $L=4096$, $T=10^6$ steps, and $N=100$.}
\label{fig:alpha_q}
\end{center}
\end{figure}

\begin{figure}
\begin{center}
\scalebox{0.7}[0.7]{\includegraphics{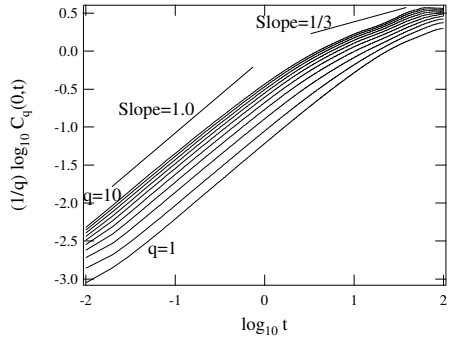}}
\caption{$\frac{1}{q}\log C_q(x',t)$ as a function of \ $\log t$ of the KPZQ model. The slope $1.0$ conflicts with  the dimensional analysis of Csah\'ok {\it et al.}. This means that the KPZ universality is destroyed. The parameters are taken as $L=1024$, $T=10^4$ steps, and $N=1$.}
\label{fig:beta_q}
\end{center}
\end{figure}

First, we compute $C_q$ of the KPZQ model. We show log-log plot of  $q$-th root of $C_q$ as a function of $x$ (Fig.\ \ref{fig:alpha_q}) and $t$ (Fig.\ \ref{fig:beta_q}). Each Figure displays $10$ curves. Each curve corresponds to $q=1,2, \ldots, 10$, increasing from the bottom to the top. Fully parallel $C_q$ curves are in small $x$ (or $t$) regime. 
 This implies the mono-affinity of the KPZQ model. The slope of curves in Fig.\ \ref{fig:alpha_q} indicates $\alpha_q = 0.75$ in short range scale. This value agrees with the theoretically predicted value $\alpha=0.75$. We use the system size $L=4096$, calculation steps $T=10^6$ steps, and ensemble number $N=100$. 
 
Similarly, curves in Fig.\ \ref{fig:beta_q} have the same slope. It means mono-growth property of the KPZQ growth. However, its value in the short range scale $\beta_q=1.0$ is not trivial. While Csah\'ok {\it et al.}, calculated the KPZ universality compatible result $\beta = 0.6$, we obtain $\beta=1.0$. We use $L=1024$, $T=10^4$ steps, and $N=1$ for this calculation. In Figs.\ \ref{fig:alpha_q} and \ref{fig:beta_q}, the KPZ scaling ($\alpha_q=0.5$ and $\beta_q=1/3$) can be slightly observed in the long range scale.  
 
\begin{figure}
\begin{center}
\scalebox{0.7}[0.7]{\includegraphics{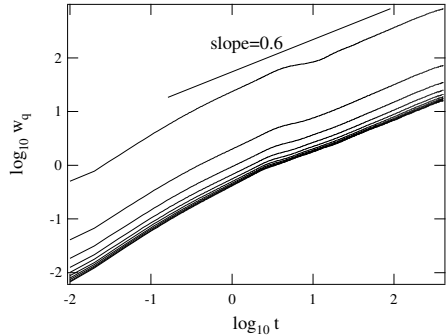}}
\caption{$\log w_q$ vs.\ $\log t$ of the KPZQ model. Although there include a little fluctuations, we can confirm the fit to the slope $0.6$. The parameters are taken as $L=1024$, $T=5\times 10^4$, and $N=200$.}
\label{fig:beta_q_direct}
\end{center}
\end{figure}

We can also evaluate $\beta_q$ using the $w_q$ scaling according to eq.\ (\ref{eq:w_q}). We show the log-log plot of $w_q$ as a function of $t$ in Fig.\ \ref{fig:beta_q_direct}. The curves correspond to $q=1,2,\ldots,10$ from the top to the bottom. The slope corresponds to $\beta_q$($\simeq 0.6$). Mono-growth property can be seen again, and the value agrees with theoretical result (the KPZ universality is recovered). 

\begin{figure}
\begin{center}
\scalebox{0.7}[0.7]{\includegraphics{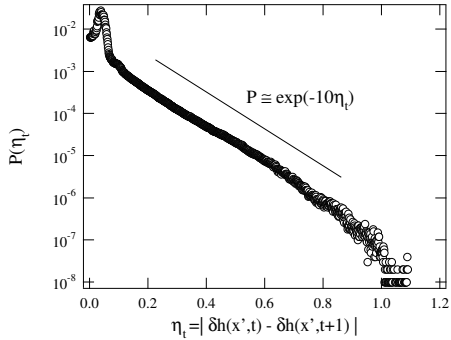}}
\caption{Distribution of neighbor-height-differences of the KPZQ model. Clear exponential part about $P(\eta_t) \sim \exp(-10\eta_t)$ can be observed. The parameters are taken as $L=1024$, $T=10^4$ steps, and $N=10$.}
\label{fig:gp_diff_kpzq}
\end{center}
\end{figure}

In addition, we introduce one more quantity in order to describe the growing rough surfaces. We consider a growth path $\delta h(x',t)$ at fixed $x'$, and define a neighbor-height-difference (or local growth velocity) $\eta_t$ as follows,
\begin{equation}
\eta_t = |  \delta h(x',t) - \delta h(x',t+1) |. \label{eq:eta_t}
\end{equation}
The size distribution of $\eta_t$ is displayed in Fig.\ \ref{fig:gp_diff_kpzq}. As can be seen in Fig.\ \ref{fig:gp_diff_kpzq}, the distribution has the exponential tail.

We have reported the results on the BDP model in our previous paper \cite{Katsuragi2}. Let me summarize the result here. The BDP model exhibits multiaffinity and multi-growth property, i.e., $\alpha_q$ and $\beta_q$ varies with $q$. The size distribution of neighbor-height-differences has the power-law tail. This power-law allows us to use the multifractal dimensional analysis method in order to obtain the asymptotic function of $\alpha_q$ and $\beta_q$ for large $q$ regime ($1/q \leq \alpha_q \leq 2/(q+1)$ and $\beta_q \simeq 1/q$). These behavior of $\alpha_q$ and $\beta_q$ are quite different from those of the KPZQ model. As mentioned before, the $\alpha_q$ and $\beta_q$ become constants in the KPZQ model. Note that the size distribution of neighbor-height-differences in the KPZQ model has the exponential tail. Therefore, we cannot apply the multifractal dimensional analysis method to the KPZQ model. We have also reported on the mono-affinity of the KPZQ model in ref.\ \ref{ref:Katsuragi3}, however, it was not performed in sufficient system size. The results could not be quantitatively reliable enough.   

As described above, the KPZQ model shows mono-affinity and the BDP model shows multiaffinity, indeed. And those relate to the distribution types of neighbor-height-differences; exponential or power-law, respectively. That is, we can distinguish the very similar profiles, ({\it e.g.}, Fig.\ \ref{fig:SurfaceProfiles}) using the multiaffine analysis.

\section{Discussion}
\label{sec:discussion}
\begin{figure}
\begin{center}
\scalebox{0.7}[0.7]{\includegraphics{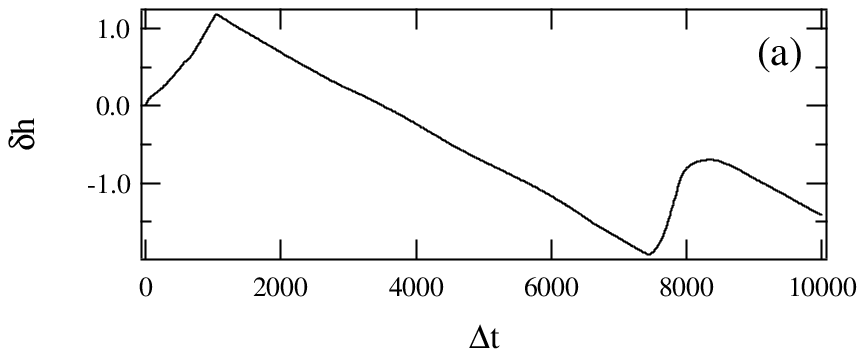}}
\scalebox{0.7}[0.7]{\includegraphics{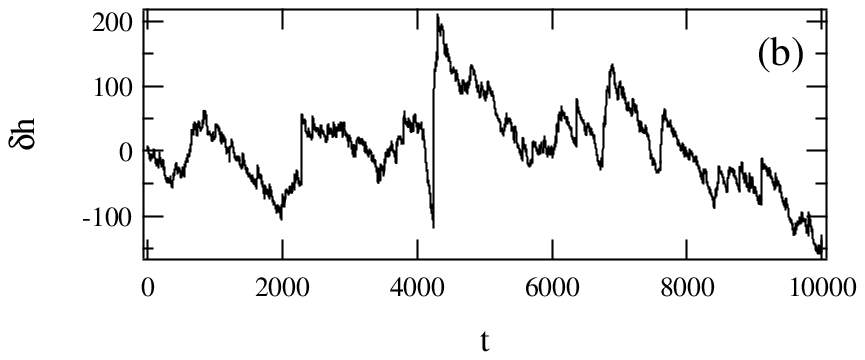}}
\caption{Typical growth path of (a) KPZQ model and (b) BDP model. The parameters are taken as $L=1024$ and $T=10^4$ steps for both models, and $\mu=3.0$ for the BDP model.}
\label{fig:GrowthProfiles}
\end{center}
\end{figure}

In the KPZQ model, $q$-th order growth exponent $\beta_q$ depends on measuring methods. The $C_q(x',t)$ based method yields $\beta_q=1.0$. Contrary, the $w_q$ method provides $\beta_q=0.6$ that concurs with theory and satisfies KPZ universality. Because $C_q(x',t)$ expresses the self-affinity of growth path, the box-count fractal dimension $D_B$ of the growth path can be written as $D_B = 2-\beta_1$ \cite{Katsuragi1,Feder1}. In the present case, we obtain $\beta_q=1.0$ for all $q$, therefore, $D_B$ becomes $1.0$. This infer that the growth path is not a rough curve, but a smooth $1$-dimensional one. We show examples of growth path of each model in Fig.\ \ref{fig:GrowthProfiles}. As expected, smooth and rough growth paths are confirmed for the KPZQ and the BDP model, respectively. 
In this mean, the value of $\beta_q$ of the BDP model does not depend on measuring methods. The growth paths of the BDP model are ordinal rough profiles \cite{Buldyrev1,Katsuragi2}.
Moreover, the scale of $\delta h$ in Fig.\ \ref{fig:GrowthProfiles}(a) is significantly smaller than that of Fig.\ \ref{fig:GrowthProfiles}(b). These data indicate that the growth of the KPZQ model is calmer than that of the BDP model, at least in the small $t$ regime.

Meakin pointed out the Hurst exponent and the roughness exponent are not always same \cite{Meakin1}. He has discriminated self-affinity and roughness scaling. His thought was limited in spatial dimension. Our result on $\beta_q$ corresponds to the discrimination between the self-affinity of growth path and the growth scaling. The $C_q$ method provides the former, and $w_q$ mehod provides the latter.
Moreover, slopes for long range scale in Figs.\ \ref{fig:alpha_q} and \ref{fig:beta_q} seem to be calm. This probably implies the asymptotic KPZ behavior ($\alpha=0.5$ and $\beta=1/3$). Simulations in larger system are needed for more detail analyses.  

Slow combustion front of paper shows multiaffinity \cite{Myllys1}. The size distribution of neighbor-height-differences (effective noise) of this system obeys power-law. These characteristics are consistent with the BDP model results. On the other hand, the profile of the Hida mountains shows mono-affinity \cite{Katsuragi1}. Furthermore, the size distribution of neighbor-height-differences of this profile shows almost exponential form. These features follow the KPZQ model results. 
However, rough surface of paper combustion seems to be caused by quenched disorder on papers. And mountain profiles probably result from the rain-fall erosion. Thus, physical mechanism of each model seems to pass each other, intuitively. Recently, Maunuksela {\it et al.}, and Myllys {\it et al.}, performed numerical simulations based on real paper noise \cite{Maunuksela1,Myllys2}. They computed the value of coefficients in the KPZ equation from experiments, and used the quenched noise measured from the real paper. They obtained mono-scaling result for this type of realistic quenched disorder simulations. We could not identify the specific origin of each noise type perfectly. More comprehensive analysis about experimental data is necessary to understand universality class in detail.
This problem is an open question. There are many other kinetic rough surface models \cite{Barabasi1}. It must be interesting to study on these models from the viewpoint of multiaffinity.

We have proved the validity of multiaffine analysis method in this paper. We believe that this analysis plays an important role to classify other scaling phenomena.  A weak point of multiaffine analysis method is that it is sensitive to fluctuations of data compared with single roughness and growth exponent analysis, particularly to calculate higher order moment (large $q$ regime).  While the method is powerful, it requires numerous data. The improvement of the method  and application to other physical systems are future problems.

\section{Concluding remarks}
We applied multiaffine analysis method to the KPZQ and the BDP model, and found some features as following. The KPZQ and the BDP models have multiaffinity and mono-affinity, respectively. The size distribution of neighbor-height-differences obeys exponential form in the KPZQ model, and obeys power-law form in the BDP model. While growth path of the BDP model is rough, that of the KPZQ model is smooth. Although the KPZQ and the BDP models cannot be distinguished only by measuring $\alpha$ and $\beta$, multiaffine analysis makes these differences clear.

\section*{Acknowledgement}
We would like to thank Professor K. Honda, Professor J. Timonen, Professor M. Merikoski, Dr. M. Myllys, Professor S. Ohta, and Professor H. Sakaguchi for their helpful suggestions, discussion, and comments.


\begin{thebibliography}{00}

\bibitem{Barabasi1} 
\label{ref:Barabasi1}
A. -L.\ Barab\'asi and H. E. Stanley: 
{\it Fractal Concepts in Surface Growth} (Cambridge University Press, Cambridge, 1995).

\bibitem{Family1} 
F. Family and T. Vicsek eds.: 
{\it Dynamics of Fractal Surfaces} (World Scientific, Singapore, 1989).

\bibitem{Kardar1} M. Kardar, G. Parisi, and Y. -C.\ Zhang: 
Phys.\ Rev.\ Lett.\ {\bf 56} (1986) 889.

\bibitem{Rubio1}
M. A. Rubio, C. A. Edwards, A. Dougherty, amd J. P. Gollub: 
Phys.\ Rev.\ Lett.\ {\bf 63} (1989) 1685. 

\bibitem{Horvath1}
V. K. Horv\'ath, F. Family, and T. Vicsek:  
J. Phys.\ A {\bf 24} (1991) L25.

\bibitem{He1}
S. He, G. L. M. K. S. Kahanda and P.-z. Wong:
Phys.\ Rev.\ Lett.\ {\bf 69} (1992) 3731.

\bibitem{Buldyrev1}
S. V. Buldyrev, A.-L. Barab\'asi, F. Caserta, S. Havlin, H. E. Stanley, and T. Vicsek: 
Phys.\ Rev.\ A {\bf 45} (1992) R8313.

\bibitem{Vicsek1}
T. Vicsek, M. Cerz\"o,and V. K. Horv\'ath: 
Physica A {\bf 167} (1990) 315.

\bibitem{Wakita1}
J. Wakita, H. Itoh, T. Matsuyama, M. Matsushita: 
J. Phys.\ Soc.\ Jpn.\ {\bf 66} (1997) 67. 

\bibitem{Zhang1}
J. Zhang, Y.-C. Zhang, P. Astr\o m, and M. T. Levinsen:  
Physica A {\bf 189} (1992) 383.

\bibitem{Myllys1} M. Myllys, J. Maunuksela, M. J. Alava, T. Ala-Nissila, and J. Timonen: 
Phys.\ Rev.\ Lett.\ {\bf 84} (2000) 1946; 
M. Myllys, J. Maunuksela, M. J. Alava, T. Ala-Nissila, J. Merikoski, and J. Timonen: 
Phys.\ Rev.\ E {\bf 64} (2001) 036101. 

\bibitem{Honjo1}
H. Honjo and S. Ohta: 
Phys.\ Rev.\ E {\bf 49} (1994) R1808.

\bibitem{Matsushita1}M. Matsushita and S. Ouchi: 
Physica D {\bf 38} (1989) 246.
  
\bibitem{Katsuragi1}
\label{ref:Katsuragi1}
H. Katsuragi and H. Honjo: 
Phys.\ Rev.\ E {\bf 59} (1999) 254.

\bibitem{Lopez1}
\label{ref:Lopez1}
 J. M. L{\'o}pez: 
Phys.\ Rev.\ Lett. {\bf 83} (1999) 4594.

\bibitem{Cshok1}
\label{ref:Cshok1}
 Z. Csah\'ok, K. Honda, and T. Vicsek: 
J.\ Phys.\ A {\bf 26} (1993) L171. 

\bibitem{Zhang2}
\label{ref:Zhang2}
 Y. -C.\ Zhang: 
J.\ Phys.\ (Paris) {\bf 51} (1990) 2129.

\bibitem{Feder1}
J. Feder: 
{\it Fractals} (Plenum, New york, 1988).  

\bibitem{Katsuragi2} 
\label{ref:Katsuragi2}
H. Katsuragi and H. Honjo:
Phys.\ Rev.\ E {\bf 67} (2003) 011601.

\bibitem{Leschhorn1}
H. Leschhorn: 
Phys.\ Rev.\ E {\bf 54} (1996) 1313.

\bibitem{Nagatani1}
T. Nagatani: 
Phys.\ Rev.\ E {\bf 58} (1998) 700.

\bibitem{Barabasi2}
A. -L.\ Barab\'asi, R. Bourbonnais, M. Jensen, J. Kert\'esz, T. Vicsek, and Y. -C.\ Zhang: 
Phys.\ Rev.\ A {\bf 45} (1992) R6951.

\bibitem{Bul2}
S. V. Buldyrev, S. Havlin, J. Kert\'esz, H. E. Stanley, and T. Vicsek: 
Phys.\ Rev.\ A {\bf 43} (1991) 7113.

\bibitem{Horvath2}
V. K. Horv\'ath, F. Family, and T. Vicsek:  
Phys.\ Rev.\ Lett.\ {\bf 67} (1991) 3207.

\bibitem{Katsuragi3}
\label{ref:Katsuragi3}
H. Katsuragi and H. Honjo: 
{\it Proc.\ of the ICCMSE 2003} (World Scientific, Singapore, 2003), p.298. 

\bibitem{Meakin1}
P. Meakin: 
{\it Fractals, scaling and growth far from equilibrium} (Canbridge Universiy Press, Cmabridge, 1998)

\bibitem{Maunuksela1} J. Maunuksela, M. Myllys, M. Merikoski, J. Timonen, T. K\"arkk\"ainen, M. S. Welling, and R. J. Wingaarden: 
Eur.\ Phys.\ J. B {\bf 33} (2003) 193.

\bibitem{Myllys2} M. Myllys, J. Maunuksela, M. Merikoski, J. Timonen, and M. Avikainen: 
Eur.\ Phys.\ J. B {\bf 36} (2003) 619. 


\end{thebibliography}
\end{document}